\documentclass[amsmath,amssymb,nofootinbib,twocolumn]{revtex4}
\usepackage{graphicx}
\usepackage{dcolumn}
\usepackage{color}
\usepackage{scrextend}
\usepackage{subfig}
\usepackage{bm}
\usepackage{amsmath,amssymb,graphicx}
\usepackage{epsfig}
\usepackage{enumerate}
\usepackage{array}
\usepackage{multirow}

\newtheorem{theorem}{Result}

\begin{document}
\title
{Landau levels of a Dirac electron in graphene\\ from non-uniform magnetic fields}
\author{Aritra Ghosh\footnote{aritraghosh500@gmail.com}}
\affiliation{School of Basic Sciences,\\ Indian Institute of Technology Bhubaneswar, Khurda, Odisha 752050, India\\
\& \\
School of Physics and Astronomy,\\ Rochester Institute of Technology, Rochester, New York 14623, USA}
\vskip-2.8cm
\date{\today}
\vskip-0.9cm

\vspace{5mm}
\begin{abstract}
The occurrence of Landau levels in quantum mechanics when a charged particle is subjected to a uniform magnetic field is well known. Considering the recent interest in the electronic properties of graphene, which admits a dispersion relation which is linear in the momentum near the Dirac points, we revisit the problem of Landau levels in the spirit of the Dirac Hamiltonian and ask if there are certain non-uniform magnetic fields which also lead to a spectrum consisting of the Landau levels. The answer, as we show, is in the affirmative. In particular, by considering isospectral deformations of the uniform magnetic field, we present explicit analytical expressions for non-uniform magnetic fields that are strictly isospectral to their uniform counterpart, thus supporting the Landau levels. 
\end{abstract}

\maketitle

\section{Introduction}
 The Dirac equation (see for example, \cite{dirac}) is a remarkable equation in physics and is one of the most important discoveries of the 20th century. While it was originally introduced in order to address the relativistic generalization of quantum mechanics, leading naturally to the notions of spin, particle-antiparticle pairs, etc., the system has attracted immense attention in the recent times, especially in the context of graphene, a two-dimensional material which is a monolayer of C atoms with the peculiarity that near the Dirac points, the electronic dispersion relation is linear in the momentum \cite{graphene0,graphene,graphene01,graphene1}. This material has intriguing properties stemming from this particular form of the dispersion relation and moreover, the absence of an energy gap implies that the low-energy physics is well described by the massless Dirac Hamiltonian \cite{graphene}. So much has been the interest in Dirac-like Hamiltonians in the context of condensed-matter physics that systems with such linear dispersion relations are now termed Dirac materials \cite{DM1,DM2,DM3,DM4,DM5}. It may also be mentioned that the Dirac Hamiltonian in two spatial dimensions has been the subject of several investigations \cite{2d1,2d2,2d3,2d4} and a key aspect of the Dirac Hamiltonian lies in its connection with quantum optics because the Hamiltonian of the so-called Dirac oscillator \cite{diracosc1,diracosc} can be mapped to that of the (anti-)Jaynes-Cummings model \cite{qpt1,diracB} describing the atomic transitions in two-level systems. 
 
 \vspace{2mm}

While considering electronic properties, an important goal is to understand how the system (particularly, the spectrum) is affected by the application of an external magnetic field. This, of course, leads us naturally to intriguing phenomena in condensed-matter physics such as diamagnetism, Hall effect, magneto-resistance, etc. It is well understood that if a uniform magnetic field is applied to a free quantum particle, the Landau levels describe the spectrum (see also, \cite{landauelec}) -- this effect is instrumental in leading to the diamagnetism due to free electrons in metals as described by Landau \cite{landau}. While the Landau levels in graphene were studied in \cite{graphene1} (see also, \cite{hall}) in the context of uniform magnetic fields, one can now ask --  are there non-uniform field configurations which may also lead to the Landau levels? Thus, one is essentially seeking isospectral deformations of the uniform magnetic field. While some progress concerning the two-dimensional Pauli Hamiltonian was reported in \cite{CNK}, in this work, we will generalize the construction to the Dirac Hamiltonian in the manner suited for graphene. The analysis is performed in both the symmetric (Coulomb) gauge and the asymmetric (Landau) gauge. It may be pointed out that some exact results on the application of magnetic fields on a Dirac electron in graphene have been reported earlier in the literature \cite{LMN,David,midya}. 

 \vspace{2mm}

The problem of finding isospectral Hamiltonians in quantum mechanics is quite old and with well-established results \cite{darboux,susy1,susy11,susy2,susy3,susy35,susy4,susy45,susyimp1,susyimp2,susy5,susy6} (see also, \cite{ratext1,ratext2,ratext3,ratext4,ratext5,ratext6,ratext7}). In fact, the classic result of Darboux \cite{darboux} says that if $\mathcal{H} \psi = \mathcal{E} \psi$ be a time-independent Schr\"odinger equation (with symbols having familiar meanings), where $\mathcal{H} = -  \frac{d^2}{dx^2} + V(x)$ (we will set $\hbar = 1$) for the potential function $V(x)$, then, if $\phi(x)$ be a particular solution of the equation with eigenvalue $\epsilon$, it turns out that for $\mathcal{E} \neq \epsilon$ with eigenfunction $\psi(x)$, the Darboux-transformed function $\tilde{\psi}(x) = W(\psi(x), \phi(x))/\phi(x)$ is a solution of $\tilde{\mathcal{H}}^\phi \tilde{\psi} = \mathcal{E} \tilde{\psi}$, where $\tilde{\mathcal{H}}^\phi  = -  \frac{d^2}{dx^2} + \tilde{V}^\phi(x)$, and 
\begin{equation}\label{darboux}
\tilde{V}^\phi(x) = V(x) - 2 \frac{d^2}{dx^2} \ln \phi(x). 
\end{equation} 
Here, $W(\psi(x), \phi(x))$ is the Wr\"onskian and $\phi(x)$ is often called the `seed' function. A special case of the Darboux transformation occurs if $\epsilon$ is the ground-state energy $\mathcal{E}_0$ of $\mathcal{H}$, in which case $\phi(x)$ must be the ground-state wavefunction, i.e., $\phi(x) = \psi_0(x)$. One then gets the framework of supersymmetric quantum mechanics in which the superpotential defines a pair of partner potentials with one of them being $V(x)$ and the other being $\tilde{V}^{\psi_0}(x)$ -- both share a common spectrum except for the ground state \cite{susy4}. Darboux transformations can also be utilized to find the so-called rationally-extended systems \cite{ratext1,ratext2,ratext3,ratext4,ratext5,ratext6,ratext7}. The specific technique that we shall apply in this study can be attributed to Abraham and Moses \cite{susy1}, and which differs from the Darboux transformation. For a suitable real constant $\lambda$, this transformation leads to new one-parameter potentials going as
\begin{equation}\label{AM}
\tilde{V}_\lambda(x) = V(x) - 2 \frac{d^2}{dx^2} \ln [J(x) + \lambda],  
\end{equation} where $J(x) = \int_{-\infty}^x \psi_0(x')^2 dx'$. These potentials are `strictly' isospectral to $V(x)$. As has been emphasized upon in \cite{susy3}, the Abraham-Moses transformation generally gives results that are distinct from the Darboux transformation; the latter generically `deletes' the eigenvalue $\epsilon$ from the spectrum associated with the potential $\tilde{V}^\phi(x)$. Nevertheless, the result (\ref{AM}) is related closely to supersymmetric quantum mechanics as it can be interpreted as a generalization of the Darboux transformation and has been detailed in \cite{susyimp1,susyimp2}. It is important to note that the parameter $\lambda$ will be taken to conform to the condition $\lambda > 0$, necessary to ensure $J(x) + \lambda > 0$ as needed for the square-integrability of the Abraham-Moses-transformed wavefunctions.

 \vspace{2mm}

The purpose of this short paper is to demonstrate that there indeed exist non-uniform magnetic fields, which, despite not being constant in space can yield the levels that are strictly isospectral to the case with a uniform magnetic field, i.e., the Landau levels. Given the current interest in graphene and Dirac/semi-Dirac materials \cite{graphene,graphene1,DM1,DM2,DM3,DM4,DM5}, such a result is expected to have interesting consequences, especially in the context of engineered magnetic barriers \cite{engg} or strain-induced pseudo-magnetic fields as observed in graphene \cite{pseudographene1,pseudographene2}.

\section{Dirac equation in magnetic field}
For graphene where the electronic properties are described by the tight-binding model, in the low-energy regime, i.e., at energies close to the Dirac points, the tight-binding Hamiltonian can be linearized to take the form of a Dirac-type (linear) Hamiltonian in two spatial dimensions (see for example, \cite{graphene}). In the presence of a gauge potential with minimal coupling, therefore, one has as the starting point, the following Hamiltonian:
\begin{equation}\label{Hmodelgeneral}
H = v_F \mathbf{\alpha} \cdot (\mathbf{p} + \mathbf{A} ),
\end{equation} where $v_F$ is the Fermi velocity while $\mathbf{A} = (A_x,A_y)$ is the vector potential with the magnetic field being defined as $B_z = \frac{\partial A_y(x,y)}{\partial x} - \frac{\partial A_x(x,y)}{\partial y}$. Notice that we have taken the electronic charge to be `$-1$' corresponding to an electron. Taking $\alpha_x = \sigma_x$ and $\alpha_y = \sigma_y$, the Dirac Hamiltonian can be expressed as
\begin{equation}\label{2dgenmatrixH}
H =
\begin{pmatrix}
0 & v_F(\mathcal{P}^* + \mathcal{A}^*(\mathbf{r})) \\
v_F(\mathcal{P} + \mathcal{A}(\mathbf{r})) & 0
\end{pmatrix} ,
\end{equation} 
where we have defined $\mathcal{P} = p_x + i p_y$ and $\mathcal{A}(\mathbf{r}) = A_x(\mathbf{r}) + i A_y (\mathbf{r})$, with $\mathbf{r} = (x,y)$. Now, taking the Dirac wavefunction to be of the form dictated by 
\begin{equation}\label{diracwave}
\Psi = \begin{pmatrix}
\psi_1 \\
\psi_2
\end{pmatrix},
\end{equation} where $\psi_{1,2} = \psi_{1,2}(x,y)$, the equation $H \Psi = E \Psi$ gives
\begin{eqnarray}
v_F(\mathcal{P}^* + \mathcal{A}^*(\mathbf{r})) \psi_2(\mathbf{r}) &=& E\psi_1(\mathbf{r}), \label{6} \\
v_F(\mathcal{P} + \mathcal{A}(\mathbf{r})) \psi_1(\mathbf{r}) &=& E \psi_2(\mathbf{r}). \label{7}
\end{eqnarray}
Combining these two, one obtains
\begin{equation}\label{KG1}
v_F^2 \big[(\mathcal{P} + \mathcal{A}(\mathbf{r})) (\mathcal{P}^* + \mathcal{A}^*(\mathbf{r}))\big] \psi_2(\mathbf{r}) = E^2 \psi_2(\mathbf{r}). 
\end{equation}
We can refer to $\psi_2$ as the wavefunction, although once it is determined, $\psi_1$ can be easily found from equation (\ref{6}). Now, for the equation (\ref{KG1}), an explicit calculation gives
\begin{eqnarray}\label{Diracgenvect}
&&\Big[ p_x^2 + p_y^2 + A_x^2 + A_y^2 + 2 (A_x p_x + A_y p_y) \nonumber \\
&&~~+ (p_xA_x) + (p_yA_y) + i (p_y A_x) -  i (p_x A_y) \Big] \psi_2(x,y) \nonumber \\
&&= \bigg( \frac{E^2 }{v_F^2} \bigg) \psi_2(x,y). 
\end{eqnarray}
The above equation is of the form $\mathcal{K} \psi_2 = \mathcal{E} \psi_2$, where $\mathcal{E} = (E/v_F)^2$ and $\mathcal{K} = p_x^2 + p_y^2 + A_x^2 + A_y^2 + 2 (A_x p_x + A_y p_y) + (p_xA_x) + (p_yA_y) + i (p_y A_x) -  i (p_x A_y)$ can be called the quasi-Hamiltonian \cite{rahul}. In what follows, we will discuss the symmetric and asymmetric gauge choices one after the other to find isospectral deformations of the uniform magnetic field.

\section{Symmetric gauge}

\subsection{Uniform magnetic field}

For a uniform magnetic field, one can choose the symmetric gauge by taking $A_x = - By$ and $A_y = Bx$, giving $B_z = 2B$. Consequently, equation (\ref{Diracgenvect}) becomes
\begin{widetext}
\begin{equation}\label{TISEtypeeqnuniform}
\bigg[- \bigg( \frac{\partial^2}{\partial x^2} + \frac{\partial^2}{\partial y^2} \bigg) + B^2 (x^2 + y^2) + 2B L_z - 2B\bigg]\psi_2(x,y) = \bigg( \frac{E^2 }{v_F^2} \bigg)\psi_2(x,y),
\end{equation}
where we have used $p_{x} = - i \frac{\partial}{\partial x}$ and $p_{y} = - i \frac{\partial}{\partial y}$. In the above equation, we have $L_z = x p_y - y p_x$ as the orbital angular momentum. Notice that the last term in the left-hand side arises as a spin-field-interaction term (`-1' orientation from the lower sign of the Pauli matrix $\sigma_z$ for $\psi_2$). It is convenient to look at this equation in the polar coordinates $(r,\theta)$, where $\left[- \left( \frac{\partial^2}{\partial x^2} + \frac{\partial^2}{\partial y^2} \right) + B^2 (x^2 + y^2) + 2B L_z - 2B\right]$ commutes with $L_z$ and a separation of variables takes place. Taking the ansatz $\psi_2(r,\theta) = e^{i m_l \theta} \rho(r)$, we will get
\begin{equation}\label{rhoeqnconstantB}
- \frac{d^2 \rho(r)}{dr^2} - \frac{1}{r} \frac{d\rho(r)}{dr} + \bigg[ B^2 r^2 + \frac{m_l^2}{r^2} + 2B m_l - 2B \bigg] \rho(r) = \bigg( \frac{E^2 }{v_F^2} \bigg)\rho(r).
\end{equation}
Defining $\chi(r) = \sqrt{r} \rho(r)$, one further gets the simplified equation
\begin{equation}\label{chi1}
- \frac{d^2 \chi(r)}{dr^2} + \bigg[ B^2 r^2 + \frac{m_l^2 - \frac{1}{4}}{r^2} \bigg] \chi(r) = \frac{[E^2 - 2B m_l v_F^2 + 2B v_F^2]}{v_F^2} \chi(r). 
\end{equation} 
The above-mentioned Schr\"odinger-like equation has the same structure as that of a particle moving in the isotonic potential \cite{iso1} and can be solved exactly to give (see \cite{iso2,isoNH,isodir} for the details) the following result:
\begin{equation}\label{constantBsolution}
\chi_{n, m_l}(r) \sim r^{|m_l| + \frac{1}{2}} e^{-\frac{Br^2}{2}} {_1}F_1(-n, |m_l| + 1, Br^2) , \quad \quad E_n^2  =   2 B v_F^2 (2n + m_l + |m_l|),
\end{equation} with $n = 0,1,2,\cdots$ and $m_l = 0, -1, -2, \cdots$. In the symmetric gauge with a uniform magnetic field, restricting to $m_l \leq 0$ for the lower spinor component $\psi_2$ is both justified and sufficient. While the radial equation admits formal solutions for $m_l > 0$, these states correspond to higher radial excitations that are degenerate with lower-$n$, negative-$m_l$ solutions. Since the energies depend only on the combination $n + \max(0, m_l)$, all distinct energy eigenvalues are already accounted for in the $m_l \leq 0$ sector. This makes the inclusion of $m_l > 0$ redundant, allowing one to construct a complete and non-redundant basis using only $m_l \leq 0$ states. Now, $_1F_1(\cdot, \cdot, \cdot)$ is the confluent hypergeometric function which can be expressed in terms of associated Laguerre polynomials $L^{|m_l|}_n(\cdot)$ by using the identity
\begin{equation}\label{identity}
{_1}F_1(-n, |m_l| + 1, Br^2) = \frac{\Gamma(n+1) \Gamma(|m_l| + 1)}{\Gamma(|m_l| + n + 1)} L^{|m_l|}_n(Br^2),
\end{equation} where $\Gamma(\cdot)$ is the Euler gamma function. The full wavefunctions are obtained as $\psi_{2,n,m_l}(r,\theta) = r^{-1/2} e^{im_l \theta} \chi_{n,m_l}(r)$ and their orthogonality is obtained directly from that of the associated Laguerre polynomials. Explicitly, one finds
\begin{equation}
\psi_{2,n,m_l}(r,\theta) = \sqrt{\frac{n! B^{|m_l| + 1}}{\pi \Gamma(n + |m_l| + 1)} } \; r^{|m_l|} e^{ - \frac{B r^2}{2} } L_n^{|m_l|}(B r^2) e^{i m_l \theta},
\end{equation}where we have also included the normalization factors. The spectrum of $E^2$ is equispaced; these are what we will refer to as the Landau levels of the Dirac system. A case which deserves special attention is $m_l = 0$ in which case the confluent hypergeometric function can be expressed in terms of the `simple' Laguerre polynomial $L^0_n(x)$. For $n = 0$, we have $L^0_0 = 1$ and this implies that the ground-state wavefunction is given by $\psi_{2,0,0}(r) \sim e^{-\frac{Br^2}{2}}$, which, unlike $\psi_{2,0,m_l}(r)$ for $m_l \neq 0$, does not vanish as $r \rightarrow 0$. This can be explained from the fact that for $m_l \neq 0$, one has $m_l^2 > 1/4$, indicating that the $1/r^2$ term in equation (\ref{chi1}) is repulsive (centered at $r = 0$) while for $m_l = 0$, the same term in equation (\ref{chi1}) becomes attractive.

\subsection{Non-uniform magnetic fields}
Concerning non-uniform magnetic fields, let us take the following profiles for the components of the vector potential (see also, \cite{CNK}): $A_x(x,y) = - By f(r)$ and $A_y(x,y) = Bx f(r)$, where $r = \sqrt{x^2 + y^2}$ is the radial variable. A straightforward calculation reveals that equation (\ref{Diracgenvect}) takes the following form:
\begin{equation}
\Big[- \nabla^2 + r^2 B^2 f(r)^2 + 2 B f(r) L_z - 2B f(r) - r B f'(r)\Big]\psi_2(r,\theta) = \bigg( \frac{E^2 }{v_F^2} \bigg) \psi_2(r,\theta).
\end{equation}
It may be mentioned that the chosen form of the vector potential is obviously not the `most' general one that leads to non-uniform magnetic fields. The reason behind our choice stems from two factors: (i) the resulting second-order equation still undergoes separation of variables in the plane-polar coordinates, (ii) putting $f(r) = 1$ immediately allows us to obtain the results for the case with a uniform magnetic field. Now, defining $\psi_{2}(r,\theta) = r^{-1/2} e^{im_l \theta} \chi(r)$, one obtains
\begin{equation}\label{chi2}
- \frac{d^2 \chi(r)}{dr^2} + \bigg[ B^2 r^2 f(r)^2 + \frac{m_l^2 - \frac{1}{4}}{r^2} + 2 B f(r) m_l - 2B f(r) - r B f'(r)\bigg] \chi(r) = \bigg( \frac{E^2 }{v_F^2} \bigg) \chi(r).
\end{equation}  \end{widetext}

We can now make use of supersymmetric factorization. It is a simple exercise to verify that for $m_l \leq 0$, the left-hand side of the above-mentioned equation can be expressed as $\mathcal{A}_{m_l}^\dagger \mathcal{A}_{m_l}$, upon defining \cite{CNK}
\begin{equation}\label{Amoperator}
\mathcal{A}_{m_l} = \frac{d}{d r} + B r f(r) - \frac{|m_l| + \frac{1}{2}}{r}. 
\end{equation}
The ground-state eigenfunction can be calculated straightforwardly; using $\mathcal{A}_{m_l} \chi_{0,m_l}(r) = 0$, an integration gives
\begin{equation}\label{GS}
\chi_{0,m_l}(r) \sim r^{|m_l| + \frac{1}{2}} e^{- \int^r B r' f(r') dr'},
\end{equation}
with the wavefunction being $\psi_{2,0,m_l}(r,\theta) = r^{-1/2} e^{i m_l \theta} \chi_{0,m_l}(r)$. We will employ the term ground-state eigenfunction to denote $\chi_{0,m_l}(r)$ whereas the ground-state wavefunction shall refer to $\psi_{2,0,m_l} = r^{-1/2} e^{im_l \theta} \chi_{0,m_l}(r)$. Note that each $m_l$ indicates a different ground-state eigenfunction as dictated by the condition $\mathcal{A}_{m_l} \chi_{0,m_l}(r) = 0$, where the operator $\mathcal{A}_{m_l}$ carries an explicit dependence on $m_l$. In the above construction, we can identify the $m_l$-dependent superpotential as
\begin{equation}
W_{m_l}(r) = B r f(r) - \frac{|m_l| + \frac{1}{2}}{r},
\end{equation} and which leads to a one-dimensional time-independent Schr\"odinger-type equation (\ref{chi2}) with the scalar potential 
\begin{eqnarray}
V_{m_l}(r) &=& B^2 r^2 f(r)^2 + \frac{m_l^2 - \frac{1}{4}}{r^2} \nonumber \\
&&~ + 2 B \bigg( f(r) m_l - f(r) - \frac{r f'(r)}{2}\bigg). \label{mdependentpotential}
\end{eqnarray}
Now, for general $f(r)$ and hence $V_{m_l}(r)$, the Abraham-Moses result (\ref{AM}) gives a family of strictly-isospectral potentials going as
\begin{equation}\label{AMVmdependent}
\tilde{V}_{m_l,\lambda}(r) = V_{m_l}(r) - 2 \frac{d^2}{dr^2} \ln [J_{m_l}(r) + \lambda], 
\end{equation} where $J_{m_l}(r) = \int_0^r \chi_{0,m_l}(r')^2 dr'$, with $\chi_{0,m_l}(r)$ being the ground-state eigenfunction (\ref{GS}) corresponding to a given $m_l$. Here, the real parameter $\lambda > 0$ parametrizes the family of strictly-isospectral potentials $\tilde{V}_{m_l,\lambda}(r)$. Using this result, a direct calculation reveals that for a given value of $m_l$ and given a function $f(r)$ characterizing the vector potential, a strictly-isospectral family $ \tilde{f}_{m_l,\lambda}(r)$ is given by (see also, \cite{CNK})
\begin{equation}\label{oneparameterf}
B \tilde{f}_{m_l,\lambda}(r) = B f(r) + \frac{1}{r} \frac{d}{d r} \ln [J_{m_l}(r) + \lambda]. 
\end{equation}
The above-mentioned expression for $ \tilde{f}_{m_l,\lambda}(r)$ is obtained by seeking its analytical form such that when substituted into the expression (\ref{mdependentpotential}), one arrives at the Abraham-Moses-transformed potentials (\ref{AMVmdependent}) after some simplification. We can now define the Abraham-Moses-transformed vector potentials as having the components $(\tilde{A}_x)_{m_l,\lambda} = - B y  \tilde{f}_{m_l,\lambda}(r)$ and $(\tilde{A}_y)_{m_l,\lambda} = B x  \tilde{f}_{m_l,\lambda}(r)$ such that the corresponding family of magnetic fields which lead to the same quantum spectrum is given by
\begin{equation}
(\tilde{B}_z)_{m_l,\lambda}(r) = B_z(r) + \frac{1}{r} \frac{d}{dr} \bigg( r \frac{d}{dr} \ln [J_{m_l}(r) + \lambda] \bigg).
\end{equation}
In the context of the two-dimensional Pauli Hamiltonian, the same result was obtained in \cite{CNK}. Our first result can now be stated as follows: 

\begin{theorem}
A two-parameter family of non-uniform magnetic fields that support the Landau levels is given by
\begin{eqnarray}
(\tilde{B}_z)_{m_l,\gamma}(r) = 2B + \frac{ r^{2|m_l|} e^{-B r^2}}{\gamma + \int_0^r {r'}^{2|m_l| + 1} e^{- B {r'}^2} dr'} \nonumber \\
+ \frac{d}{dr} \bigg( \frac{ r^{2|m_l|+1} e^{-B r^2}}{\gamma + \int_0^r {r'}^{2|m_l| + 1} e^{- B {r'}^2} dr'} \bigg),\label{mainresult}
\end{eqnarray} where $\gamma$ is a suitable positive constant and other symbols have their usual meanings. 
\end{theorem}

{\it Proof --} As we are interested in the Landau quantization, we will take $f(r) = 1$ which gives $B_z = 2B$, corresponding to the case of equation (\ref{chi1}). Since $f(r) = 1$, the ground-state eigenfunction $\chi_{0,m_l}(r)$ is given from (\ref{GS}) as
\begin{equation}\label{GS1}
\chi_{0,m_l}(r) = N_{m_l} r^{|m_l| + \frac{1}{2}} e^{- \frac{B r^2}{2}},
\end{equation} where $N_{m_l}$ is a normalization factor. This matches with the $n=0$ case of the result (\ref{constantBsolution}). From the result (\ref{oneparameterf}), it then immediately follows that
\begin{equation}
\tilde{f}_{m_l,\lambda}(r) = 1 + \bigg(\frac{1}{B}\bigg) \frac{N_{m_l}^2 r^{2|m_l|} e^{-Br^2}}{\lambda + N_{m_l}^2 \int_0^r {r'}^{2|m_l| + 1} e^{- B {r'}^2} dr'},
\end{equation} where the subscript $m_l$ makes it explicit that the result is obtained for a particular $m_l$. From this follows non-uniform magnetic fields admitting the Landau levels (for a chosen $m_l$). Note that the energy levels are not equispaced and the largest energy separation is between $n=0$ and $n=1$, allowing one to observe the quantum Hall effect in graphene even at room temperature \cite{hall}. A direct calculation verifies the result (\ref{mainresult}) with $\gamma = \lambda/N_{m_l}^2$.\\

\begin{figure}
\begin{center}
\includegraphics[scale=1]{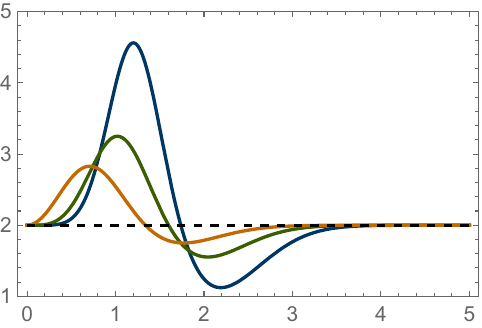}
\caption{Spatial profile of $(\tilde{B}_z)_{m_l,\gamma}(r)$ which isospectral to the uniform field $B_z = 2B$, as a function of $r$ for $B=1$, $\gamma = 1$, upon taking $m_l = -1$ (orange), $m_l = -2$ (green), and $m_l = -3$ (blue). The black-dashed line represents the uniform field $B_z = 2B$.}
\label{fig}
\end{center}
\end{figure}

Furthermore, we have the result
\begin{eqnarray}
&&\int_0^r {r'}^{2|m_l| + 1} e^{- B {r'}^2} dr' \\
&&= \frac{r^{2|m_l|}}{2B (B r^2)^{|m_l|}} \big[ \Gamma(|m_l|+1) - \Gamma(|m_l| + 1, Br^2) \big], \nonumber
\end{eqnarray} where $\Gamma(\cdot)$ is the Euler gamma function which appeared earlier in the identity (\ref{identity}) while $\Gamma(\cdot, \cdot)$ is the upper incomplete gamma function. Using this, an analytic evaluation of the non-uniform magnetic fields (\ref{mainresult}) is possible. The variation of the magnetic field (\ref{mainresult}) has been illustrated in Fig. (\ref{fig}) which clearly demonstrates its spatial profile. For large $r$, it converges to the uniform field $B_z = 2B$ and the smaller the value of $|m_l|$, the faster this convergence is. \\

It must be emphasized that the operator $\mathcal{A}_{m_l}$ and its conjugate involve the quantum number $m_l$. What this means is that the spectrum of $\mathcal{A}_{m_l}^\dagger \mathcal{A}_{m_l}$ is labeled by the quantum number $n$ but for fixed $m_l$. While this indicates that we have only obtained the partial spectrum (for a given $m_l$), it is remarkable to note that for $m_l \leq 0$, the spectrum (\ref{constantBsolution}) due to a uniform magnetic field $2B$ is $E_n^2 = 4B n v_F^2$, independent of $m_l$.  Nevertheless, for each $\mathcal{A}_{m_l}$, equation (\ref{Amoperator}) defines an $m_l$-dependent ground-state eigenfunction and the non-uniform magnetic fields (\ref{mainresult}) are $m_l$-dependent. As shown in \cite{CNK}, the above-mentioned non-uniform magnetic fields have the same magnetic flux as their uniform counterpart.

\section{Asymmetric (Landau) gauge}\label{asymsec}
Let us take the asymmetric gauge in which $A_x = 0$ and $A_y = A_y(x)$, giving $B_z = \frac{\partial A_y(x)}{\partial x}$. Due to the translation symmetry in the $y$-direction, taking the Dirac wavefunction to be of the form
\begin{equation}\label{diracwave2}
\Psi = e^{iky} \begin{pmatrix}
\psi_1 \\
i \psi_2
\end{pmatrix},
\end{equation} where $\psi_{1,2} = \psi_{1,2}(x)$, the Dirac equation $H \Psi = E \Psi$ becomes
\begin{eqnarray}
v_F\bigg[ \frac{d}{d x} + k + A_y(x) \bigg] \psi_2(x) &=& E\psi_1(x),  \\
v_F\bigg[- \frac{d}{d x}  + k  + A_y(x) \bigg] \psi_1(x) &=& E \psi_2(x). 
\end{eqnarray}
Combining these, one gets for $\psi_2$, the following equation: 
\begin{equation}\label{KG22}
\bigg[ - \frac{d^2}{dx^2} + [k + A_y(x)]^2 - \frac{dA_y(x)}{dx}\bigg] \psi_2(x)= \bigg(\frac{E^2}{v_F^2}\bigg) \psi_2(x),
\end{equation}
the left-hand side of which can be factorized as $\mathcal{A}^\dagger_k \mathcal{A}_k$, where
\begin{equation}
\mathcal{A}_k =  \frac{d}{d x} + k + A_y(x). 
\end{equation}
The ground state is obtained as $\mathcal{A}_k \psi_{2,0,k} = 0$, formally giving
\begin{equation}\label{GSpsi20exp}
\psi_{2,0,k}(x) \sim e^{- \int^x (k + A_y(x')) dx'},
\end{equation} where the subscript $k$ indicates the $k$-dependence of the wavefunction via the explicit dependence of $\mathcal{A}_k$ on $k$. Equation (\ref{KG22}) can be interpreted as a time-independent Schr\"odinger-type equation with the effective potential
\begin{equation}\label{Vxmag}
V_k(x) = [k + A_y(x)]^2 - \frac{dA_y(x)}{dx}.
\end{equation}
We can now discuss the Landau levels. 

\subsection{Landau levels}
Let us take $A_y(x) = Bx$. This implies $B_z = B$ and equation (\ref{KG22}) becomes
\begin{equation}
\bigg[ - \frac{d^2}{dx^2} + B^2\bigg(x + \frac{k}{B}\bigg)^2\bigg] \psi_2(x)= \bigg( \frac{E^2}{v_F^2} + B \bigg)\psi_2(x),
\end{equation}
which has the same structure as the time-independent Schr\"odinger equation for a harmonic oscillator. Thus, it can be solved to give
\begin{equation}\label{solutionxlandau}
\psi_{2,n,k}(x) \sim e^{-\frac{B (x + k/B)^2}{2}} H_n \bigg(\sqrt{B}x + \frac{k}{\sqrt{B}}\bigg), \quad E_n^2 = 2Bv_F^2 n, 
\end{equation} where $n = 0,1,2,\cdots$ and $k$ can vary continuously. This gives the Landau levels $E_n = \pm \sqrt{2Bn} v_F$. 

\subsection{Landau levels from non-uniform fields}
One can now present a family of strictly-isopectral magnetic fields which constitutes our second result. The following is true:

\begin{theorem}
A two-parameter family of non-uniform magnetic fields that support the Landau levels is given by
\begin{equation}
(\tilde{B}_z)_{k,\delta}(x) = B + \frac{d}{dx} \Bigg( \frac{ e^{-B \big(x + \frac{k}{B}\big)^2}}{\delta + \int_{-\infty}^x  e^{-B \big(x' + \frac{k}{B}\big)^2} dx'} \Bigg), \label{mainresult2}
\end{equation} where $\delta$ is a suitable positive constant and other symbols have their usual meanings. 
\end{theorem}

\begin{figure}
\begin{center}
\includegraphics[scale=0.55]{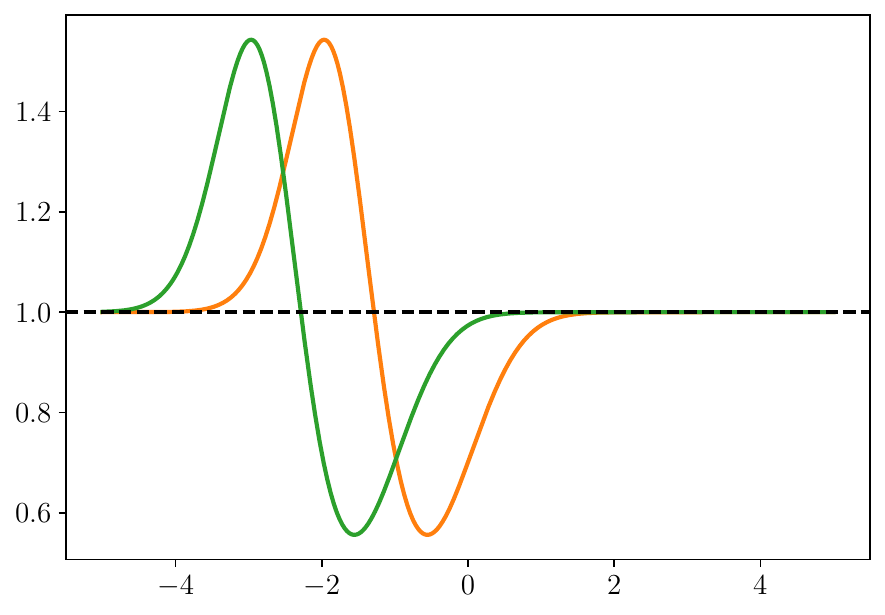}
\caption{Spatial profile of $(\tilde{B}_z)_{k,\delta}(x)$ which isospectral to the uniform field $B_z = B$, as a function of $x$ for $B=1$, $\delta = 1$, upon taking $k = 1$ (orange) and $k = 2$ (green). The black-dashed line represents the uniform field $B_z = B$.}
\label{fig2}
\end{center}
\end{figure}

{\it Proof --} Resorting to the Abraham-Moses transformation (\ref{AM}) and starting with the potential (\ref{Vxmag}), one can get a family of strictly-isospectral potentials as
\begin{equation}
\tilde{V}_{k,\lambda}(x) = V_k(x) - 2 \frac{d^2}{dx^2} \ln [J_k(x) + \lambda], 
\end{equation}
where $J_k(x) = \int_{-\infty}^x \psi_{2,0,k}(x')^2 dx'$. Invoking equation (\ref{Vxmag}), a direct calculation reveals that one gets a family of vector potentials going as
\begin{equation}
(\tilde{A}_y)_{k,\lambda}(x) = A_y(x) + \frac{d}{dx} \ln [J_k(x) + \lambda].
\end{equation}
These potentials define a family of strictly-isospectral magnetic fields. Keeping in mind the problem of Landau levels and therefore taking $A_y(x) = B x$, we can finally get our result (\ref{mainresult2}) upon putting $\psi_{2,0,k}(x) = N_k  e^{-\frac{B (x + k/B)^2}{2}}$ for a normalizing factor $N_k$ by expressing $\delta = \lambda/N_k^2$. The variation of the magnetic field (\ref{mainresult2}) has been illustrated in Fig. (\ref{fig2}) which clearly shows its spatial profile. For large $|x|$, it converges to the uniform field $B_z = B$.

\section{Discussion}
The appearance of Landau levels when a charged quantum particle is applied with a uniform magnetic field is well known. In the preceding analysis, we described how one can obtain non-uniform magnetic fields which are strictly isospectral to the case with a uniform magnetic field. To summarize, in the symmetric gauge, it was convenient to introduce the polar coordinates $(r,\theta)$ and a further transformation of the radial wavefunction $\rho(r) = r^{-1/2} \chi(r)$ which maps the problem to an `effective' equation (\ref{chi2}) which has the desired spectrum for $E^2$; on this we applied supersymmetric factorization and the Abraham-Moses transformation (\ref{AM}). Thus, the so-called wavefunctions $\chi_{n,m_l}(r)$ with ground-state normalizing factors $N_{m_l}$ are distinct from the true wavefunctions $\psi_{2,n,m_l}(r,\theta) = r^{-1/2} e^{i m_l \theta} \chi_{n,m_l}(r)$. Focusing on $m_l \leq 0$, for each $m_l = 0, -1, -2, \cdots$, we found distinct non-uniform magnetic fields (\ref{mainresult}) supporting the Landau levels. The case with $m_l = 0$ is special because the ground-state wavefunction does not vanish as $r \rightarrow 0$, as discussed in the paragraph below the identity (\ref{identity}). It is easy to see from the result (\ref{mainresult}) that while for $m_l = -1,-2,\cdots$, $\lim_{r \rightarrow 0} (\tilde{B}_z)_{m_l < 0,\gamma}(r) = \lim_{r \rightarrow \infty} (\tilde{B}_z)_{m_l < 0,\gamma}(r) = 2B$, we have for $m_l = 0$, $\lim_{r \rightarrow \infty} (\tilde{B}_z)_{m_l = 0,\gamma}(r) = 2B$ but $\lim_{r \rightarrow 0} (\tilde{B}_z)_{m_l = 0,\gamma}(r) = 2 (B + \gamma^{-1})$, an observation also pointed out in \cite{CNK}. 
 \vspace{2mm}

Now, looking at the results from the asymmetric gauge, it is easy to observe that they turn out to be $k$-dependent, where $k$ is the momentum quantum number (continuous) in the $y$-direction. While $k$ does not impact the spectrum, it does show up in the ground-state wavefunction (\ref{GSpsi20exp}) and hence, $k$ parametrizes the family of non-uniform magnetic fields (\ref{mainresult2}) supporting the Landau levels. We must note that the spectrum of $E^2$-levels is the same in both the cases (as anticipated) since for $m_l \leq 0$, we have from (\ref{constantBsolution}), the levels $E_n^2 = 4B n v_F^2$ for the (uniform) magnetic field $B_z = 2B$ which is equivalent to $E_n^2 = 2B n v_F^2$ obtained in (\ref{solutionxlandau}) for the (uniform) magnetic field $B_z = B$. Notably, our analysis reveals two-parameter families of magnetic fields as compared to one-parameter families of potentials in the original form of the Abraham-Moses result (\ref{AM}), with the emergence of $m_l$ or $k$ due to reducing the two-dimensional problem to a one-dimensional one. It is important to mention that while the Landau levels of the original problem with a uniform magnetic field admit a strong degeneracy due to $m_l$ or $k$, this degeneracy gets lifted in the case of the non-uniform magnetic fields found by invoking the Abraham-Moses transformation as these are for fixed values of $m_l$ or $k$; for the symmetric gauge, the same was emphasized upon in \cite{CNK} considering the two-dimensional Pauli Hamiltonian. It may be noted that following \cite{susy45}, one can find multi-parameter families of non-uniform magnetic fields that support the Landau levels; however, we will not pursue this here. An important question that arises out of the present investigation is -- how does one practically generate non-uniform magnetic fields that depend on the quantum number $m_l$ or $k$? We shall keep this issue to be addressed in future works.

 \vspace{2mm}

Finally, since we are discussing the Dirac Hamiltonian in magnetic fields in the light of supersymmetric quantum mechanics, let us conclude this study by pointing out the clear connection with Rabi oscillations which has been indicated by some authors earlier (see for example, \cite{rabi1,rabi2}), but now within the premise of supersymmetric factorization. In the case of the uniform magnetic field, using the symmetric choice $A_x = - By$ and $A_y = Bx$, leading to $B_z = 2B$, the Hamiltonian (\ref{Hmodelgeneral}) can be expressed as
\begin{equation}\label{Hamunifinal}
H = v_F \big[ \sigma_x (p_x - By) +  \sigma_y (p_y + Bx)\big].
\end{equation}
Defining $\pi_x = p_x - By$ and $\pi_y = p_y + Bx$, we can define the ladder operators for the Landau levels as
\begin{equation}
a = \frac{1}{\sqrt{4B}} \left( \pi_x - i \pi_y \right), \quad \quad a^\dagger = \frac{1}{\sqrt{4B}} \left( \pi_x + i \pi_y \right),
\end{equation} which satisfy the bosonic commutation algebra $[a,a^\dagger] = 1$. Using these and also the spin operators $\sigma_\pm = \frac{1}{2} (\sigma_x \pm i \sigma_y)$, the Hamiltonian (\ref{Hamunifinal}) can be expressed as
\begin{equation}
H = v_F \sqrt{4B} \big( \sigma_+ a + \sigma_- a^\dagger \big),
\end{equation} which is the Jaynes-Cummings model with coupling $g = v_F \sqrt{4B}$. This points out the connection with Rabi oscillations with resonant excitation exchange \cite{JC_book}. However, one can also obtain the anti-Jaynes-Cummings model under the transformation $B \rightarrow -B$ which preserves the spectrum but flips the chirality, i.e., the spinor structure, leading to a physical picture of Rabi oscillations with counter-rotating-like excitation exchange \cite{JC_book}. 

\vspace{2mm}

Since under $B \rightarrow -B$, we get $a \leftrightarrow a^\dagger$, the supersymmetric relationship between the Jaynes-Cummings and anti-Jaynes-Cummings models is transparent (see also, \cite{susyJC}). As the Dirac oscillator \cite{diracosc} can be mapped to the anti-Jaynes-Cummings model \cite{qpt1,diracB}, it then only seems natural that the Hamiltonian (\ref{Hamunifinal}) should be expressible in the form of a Dirac oscillator. Indeed this is true and a simple calculation reveals that an equivalent description is $H = v_F \big[ \sigma_x ( p_x + i \sigma_z B x ) + \sigma_y ( p_y + i \sigma_z B y ) \big]$, which is just the $(2+1)$-dimensional Dirac oscillator \cite{diracosc} (massless) with frequency $\omega = -B$. The problem of Landau levels is thus isospectral to the Dirac oscillator whose frequency has been appropriately chosen (see also, \cite{diracB}).\\

\textbf{Acknowledgements:} The author is grateful to C. Nagaraja Kumar for insightful discussions and is also thankful to the Department of Physics, Panjab University for hospitality where this work was initiated. Useful correspondences with Pavel Exner, Ariel Edery, Akash Sinha, Bijan Bagchi, and Bhabani Prasad Mandal are gratefully acknowledged. The author is thankful to Akash Sinha for help in preparing the figures.


\begin{thebibliography}{99}
%%%%%%%%%%%%%%%%%%%%%%%%%%%%%%%%

\bibitem{dirac}
L. I. Schiff, {\it Quantum Mechanics}, McGraw-Hill (New York) (1955).

\bibitem{graphene0}
P. R. Wallace, Phys. Rev. \textbf{71}, 622 (1947). 

\bibitem{graphene}
A. H. Castro Neto, F. Guinea, N. M. R. Peres, K. S. Novoselov, and A. K. Geim, Rev. Mod. Phys. \textbf{81}, 109 (2009).

\bibitem{graphene01}
M. Kamfor, S. Dusuel, K. P. Schmidt, and J. Vidal, Phys. Rev. B \textbf{84}, 153404 (2011).

\bibitem{graphene1}
M. O. Goerbig, Rev. Mod. Phys. \textbf{83}, 1193 (2011).

\bibitem{DM1}
T. O. Wehling, A. M. Black-Schaffer, and A. V. Balatsky, Adv. Phys. \textbf{63}, 1 (2014). 

\bibitem{DM2}
J. Cayssol, C. R. Phys. \textbf{14}, 760 (2013). 

\bibitem{DM3}
C. Zhong, Y. Chen, Y. Xie, Y.-Y. Sun, and S. Zhang, Phys. Chem. Chem. Phys. \textbf{19}, 3820 (2017). 

\bibitem{DM4}
M. D. Uryszek, E. Christou, A. Jaefari, F. Kr\"uger, and B. Uchoa, Phys. Rev. B \textbf{100}, 155101 (2019).

\bibitem{DM5}
D. Asafov and I. Pavlov, Phys. Rev. B \textbf{110}, 125126 (2024). 

\bibitem{2d1} 
S.-H. Dong, X.-W. Hou, and Z.-Q. Ma, Phys. Rev. A \textbf{58}, 2160 (1998).

\bibitem{2d2} 
S.-H. Dong and Z.-Q. Ma, Phys. Lett. A \textbf{312}, 78 (2003). 
 
\bibitem{2d3}
Y. Sucu and N. \"Unal, J. Math. Phys. \textbf{48}, 052503 (2007).
 
\bibitem{2d4} 
L. Menculini, O. Panella, and P. Roy, Phys. Rev. D \textbf{87}, 065017 (2013).

\bibitem{diracosc1} 
D. It\^{o}, K. Mori, and E. Carriere, Nuovo Cimento A {\bf 51}, 1119 (1967).

\bibitem{diracosc}
M. Moshinsky and A. Szczepaniak, J. Phys. A: Math. Gen. \textbf{22}, L817 (1989).

\bibitem{qpt1}
A. Bermudez, M. A. Martin-Delgado, and E. Solano, Phys. Rev. A {\bf  76}, 041801(R) (2007). 

\bibitem{diracB}
B. P. Mandal and S. Verma, Phys. Lett. A \textbf{374}, 1021 (2010).

\bibitem{landauelec}
A. Edery and Y. Audin, J. Phys. Commun. \textbf{3}, 025013 (2019).

\bibitem{landau}
L. D. Landau, Z. Phys. \textbf{64}, 629 (1930).

\bibitem{hall}
Y. Zheng and T. Ando, Phys. Rev. B \textbf{65}, 245420 (2002).

\bibitem{CNK}
A. Khare and C. N. Kumar, Mod. Phys. Lett. A \textbf{8}, 523 (1993).

\bibitem{LMN}
\c{S}. Kuru, J. Negro, and L. M. Nieto, J. Phys.: Condens. Matter \textbf{21}, 455305 (2009). 

\bibitem{midya}
B. Midya and D. J. Fern\'andez, J. Phys. A: Math. Theor. \textbf{47}, 285302 (2014).

\bibitem{David}
M. Castillo-Celeita and D. J. Fern\'andez C., J. Phys. A: Math. Theor. \textbf{53}, 035302 (2020).

\bibitem{darboux}
G. Darboux, C. R. Acad. Sci. (Paris) \textbf{94}, 1456 (1882).

\bibitem{susy1}
P. B. Abraham and H. E. Moses, Phys. Rev. A \textbf{22}, 1333 (1980).

\bibitem{susy11}
B. Mielnik, J. Math. Phys. \textbf{25}, 3387 (1984).

\bibitem{susy2}
M. M. Nieto, Phys. Lett. B \textbf{145}, 208 (1984).

\bibitem{susy3}
M. Luban and D. L. Pursey, Phys. Rev. D \textbf{33}, 431 (1986).

\bibitem{susy35}
R. W. Haymaker and A. R. P. Rau, Am. J. Phys. \textbf{54}, 928 (1986).

\bibitem{susy4}
A. Khare and U. Sukhatme, J. Phys. A: Math. Gen. \textbf{22}, 2847 (1989).

\bibitem{susy45}
W.-Y. Keung, U. P. Sukhatme, Q. Wang, and T. D. Imbo, J. Phys. A: Math. Gen. \textbf{22}, L987 (1989). 

\bibitem{susyimp1}
F. Cooper, A. Khare, and U. Sukhatme, Phys. Rep. \textbf{251}, 267 (1995).

\bibitem{susyimp2}
G. Junker, {\it Supersymmetric Methods in Quantum and Statistical Physics}, Springer-Verlag (Berlin, Heidelberg) (1996).

\bibitem{susy5}
B. Bagchi and R. Roychoudhury, J. Phys. A: Math. Gen. \textbf{33}, L1 (2000).

\bibitem{susy6}
A. R. P. Rau, J. Phys. A: Math. Gen. \textbf{37}, 10421 (2004).

\bibitem{ratext1}
D. G\'omez-Ullate, N. Kamran, and R. Milson, J. Phys. A: Math. Gen. \textbf{37}, 10065 (2004).

\bibitem{ratext2} 
C. Quesne, J. Phys. A: Math. Theor. \textbf{41}, 392001 (2008).

\bibitem{ratext3} 
J. F. Cari\~nena, A. M. Perelomov, M. F. Ra\~nada, and M. Santander, J. Phys. A: Math. Theor. \textbf{41}, 085301 (2008).

\bibitem{ratext4}
B. Bagchi and C. Quesne, J. Phys. A: Math. Theor. \textbf{43}, 305301 (2010).

\bibitem{ratext5}
Y. Grandati, Ann. Phys. (N.Y.) \textbf{326}, 2074 (2011). 

\bibitem{ratext6}
D. G\'omez-Ullate, Y. Grandati, and R. Milson, J. Phys. A: Math. Theor. \textbf{47}, 015203 (2014).

\bibitem{ratext7}
R. K. Yadav, S. Banerjee, N. Kumari, A. Khare, and B. P. Mandal, Ann. Phys. (N.Y.) \textbf{436}, 168679 (2022). 

\bibitem{engg}
A. De Martino, L. Dell’Anna, and R. Egger, Phys. Rev. Lett. \textbf{98}, 066802 (2007).

\bibitem{pseudographene1}
S. Zhu, Y. Huang, N. N. Klimov, D. B. Newell, N. B. Zhitenev, J. A. Stroscio, S. D. Solares, and T. Li, Phys. Rev. B \textbf{90}, 075426 (2014).

\bibitem{pseudographene2}
D.-H. Kang, H. Sun, M. Luo, K. Lu, M. Chen, Y. Kim, Y. Jung, X. Gao, S. J. Parluhutan, J. Ge, S. W. Koh, D. Giovanni, T. C. Sum, Q. J. Wang, H. Li, and D. Nam, Nat. Commun. \textbf{12}, 5087 (2021).

\bibitem{rahul}
B. Bagchi and R. Ghosh, J. Math. Phys. \textbf{62}, 072101 (2021).

\bibitem{iso1}
Y. Weissman and J. Jortner, Phys. Lett. A \textbf{70}, 177 (1979).

\bibitem{iso2}
I. I. Gol'dman and V. D. Krivchenkov, {\it Problems in Quantum Mechanics}, Pergamon Press
(London) (1961).

\bibitem{isoNH}
A. Ghosh and A. Sinha, J. Phys.: Conf. Ser. \textbf{2986}, 012004 (2025).

\bibitem{isodir}
A. Ghosh and B. P. Mandal, Phys. Lett. A \textbf{545}, 130488 (2025).

\bibitem{rabi1}
T. M. Rusin and W. Zawadzki, Phys. Rev. B \textbf{78}, 125419 (2008).

\bibitem{rabi2}
R. Guti\'errez-J\'auregui and H. J. Carmichael, Phys. Scr. \textbf{93}, 104001 (2018).

\bibitem{JC_book}
L. Allen and J. H. Eberly, {\it Optical Resonance and Two-Level Atoms}, Dover Publications (New York) (1987).

\bibitem{susyJC}
I. A. Bocanegra-Garay, M. Castillo-Celeita, J. Negro, L. M. Nieto, and F. J. G\'omez-Ruiz, Phys. Rev. Research \textbf{6}, 043218 (2024).


\end{thebibliography}
\end{document}